\documentclass[twocolumn,showpacs]{revtex4}
\def\be{\begin{equation}}
\def\ee{\end{equation}}

\def\pp{\psi(2S)}
\def\jp{J/\psi}

\begin{document}
\title{Hadro-Charmonium}

\author{S. Dubynskiy$^{1}$ and M.B. Voloshin$^{2,3}$}

\affiliation{
$^1$School of Physics and Astronomy, University of Minnesota, Minneapolis, MN
55455, USA\\
$^2$William I. Fine Theoretical Physics Institute, University of
Minnesota, Minneapolis, MN 55455, USA \\
$^3$Institute of Theoretical and Experimental Physics, Moscow 117259, Russia}

\begin{abstract}
We argue that relatively compact charmonium states, $\jp$, $\pp$, $\chi_c$, can very likely be bound inside light hadronic matter, in particular inside higher resonances made from light quarks and/or gluons. The charmonium state in such binding essentially retains its properties, so that the bound system decays into light mesons and the particular charmonium resonance. Thus such bound states of a new type, which we call hadro-charmonium, may explain the properties of
some of the recently observed resonant peaks, in particular of $Y(4.26)$, $Y(4.32-4.36)$, $Y(4.66)$, and $Z(4.43)$. We discuss further possible implications of the suggested picture for the observed states and existence of other states of hadro-charmonium and hadro-bottomonium.  
\end{abstract}

\maketitle

Recent experimental observations near the charm threshold strongly suggest that the spectrum of resonances with hidden charm is remarkably more rich than suggested by the standard quark-antiquark template and very likely includes states where the heavy-quark $c \bar c$ pair is accompanied by light quarks and/or gluons. In particular, the peak $X(3872)$ originally observed\cite{bellex} through its decay into $\pi^+ \pi^- \jp$ is most likely caused by the near-threshold `molecular'\cite{ov,xmany} dynamics of an $S$ wave pair of charmed mesons: $D^0 {\bar D}^{*0} + \bar D^0 D^{*0}$ (recent reviews of this subject can be found in Refs.~\cite{egmr,mvc,go}). Some of the other recently found states, those with mass around 3940\,MeV, can be understood in terms of higher levels of conventional $c \bar c$ charmonium\cite{egmr,mvc}. However at least four of the observed peaks, namely the $J^{PC}=1^{--}$ resonances, $Y(4.26)$, the `broad structure' at 4.32\,GeV~\cite{babar432} or/and 4.36\,GeV~\cite{belle432}: $Y(4.32-4.36)$, another peak $Y(4.66)$, and the unquestionably exotic peak $Z(4.43)$, appear to display properties that hint at that these states contain a particular charmonium resonance, $J/\psi$, or $\pp$, that stays essentially intact inside a more complex hadronic structure\cite{mvc}. The only observed decay channels for these states are those containing a specific charmonium resonance: $\jp$ for the $Y(4.26)$ and $\pp$ for the rest, accompanied by light mesons, pions or Kaons. Any decays into states with a different `non preferred' charmonium resonance (e.g. $\pp$ instead of $\jp$) as well as decays into pairs of $D$ mesons, possibly accompanied by pions, have not been observed so far, and at least in some instances significant upper bounds on such decays do exist\cite{noother}.

The discussed here hadronic objects, containing a particular charmonium state embedded in a light hadronic matter, can be called hadro-charmonium. We believe that it is appropriate and helpful to distinguish such objects from `molecular' states as well as from more generic multiquark ones. The molecular states, like $X(3872)$, contain pairs of heavy-light mesons that largely do not overlap within the `molecule', while possible more tightly bound states where the heavy and light quarks and antiquarks (quasi)randomly overlap would be generic multiquark states. In hadro-charmonium the heavy and light degrees of freedom are separated by their size scale rather than spatially: a compact colorless charmonium sits inside a larger `blob' of light hadronic matter~\cite{ghm}. The interaction between colorless charmonium `kernel' and the light `shell' is through a QCD analog of the van der Waals force. We argue that for a relatively compact charmonium state the interaction may be sufficiently weak to preserve its particular quantum state, i.e. to prevent a breakup of the charmonium or a strong mixing with other charmonium states. On the other hand, we argue that the same interaction is sufficiently strong to form a bound system, at least in the case where the light hadronic state is that of a highly excited resonance. The observation of the $Y$ peaks and of the $Z(4.43)$ suggests that the binding in fact occurs at a moderately low excitation of the light hadronic matter. Furthermore, a formation of hadro-charmonium is favored for higher charmonium resonances $\pp$, $\chi_c$,  as compared to the lowest states $\jp$ and $\eta_c$. Thus one would expect the spectrum of hadro-charmonium resonances decaying into $\pp$ and light mesons to be more prolific than that of the states decaying into channels with $\jp$. Moreover, we fully expect existence of resonances decaying into $\chi_{cJ}$ and one or more pions, in the same mass range as the discussed $Y$ and $Z$ peaks.

It can be noted that the binding of specific charmonium resonances in hadronic matter due to the van der Waals type force has been discussed in the literature since some time ago\cite{bst} in connection with bound states of  $\jp$ and $\pp$ in nuclei. Such binding necessarily occurs in sufficiently large nuclei, although the status of binding to light nuclei and in particular to a single nucleon is still unclear\cite{sv}. In this paper we essentially extend the arguments of Ref.\cite{sv} to the case when a spatially `large' state of light hadronic matter is an excited meson rather than a heavy nucleus, which allows to conclude that at least some levels of charmonium very likely get bound inside such a meson. 

The interaction of a compact charmonium state (call it generically $\psi$) with long wave length gluonic field inside light hadronic matter can be described in terms of the multipole expansion in QCD\cite{gottfried,mv79,peskin} with the leading term being the E1 interaction with the chromo-electric field ${\vec E}^a$. The effective Hamiltonian arising in the second order in this interaction can then be written as
\be
H_{\rm eff}= -{1 \over 2} \, \alpha^{(\psi)} \, E_i^a E_i^a~,
\label{hdiag}
\ee
where $\alpha^{(\psi)}$ is the (diagonal) chromo-electric polarizability of the state $\psi$ having the dimension of volume, and which can be expressed in terms of the Green's function ${\cal G}$ of the heavy quark pair in a color octet state:
\be
\alpha^{(\psi)} = {1 \over 16} \langle \psi | \xi^a r_i {\cal G} r_i \xi^a | \psi \rangle~,
\label{alpsi}
\ee
with ${\vec r}$ being the relative position of the quark and the antiquark and $\xi^a$ the difference of the color generators acting on them: $\xi^a = t_c^a- t_{\bar c}^a$ (a detailed discussion can be found e.g. in the review \cite{mvc}). Finally, the QCD coupling $g$ is included in the normalization of the field strength in Eq.(\ref{hdiag}), so that e.g. the gluon Lagrangian in this normalization takes the form $-(1/4 g^2) F^2$.

The diagonal chromo-electric polarizability is not yet known for either of the charmonium states, although it can be directly measured for the $\jp$ resonance\cite{mvpol}. What is known is the off-diagonal chromo-polarizability $\alpha^{(\psi \psi')}$ describing the strength of the amplitude of the transition $\pp \to \pi^+ \pi^- \jp$: $|\alpha^{(\psi \psi')}| \approx 2 \,$ GeV$^{-3}$~\cite{mvc,sv}. On general grounds, the diagonal chromo-polarizability for each of the $\jp$, $\pp$ and $\chi_c$ states should be real and positive, and for the former two the Schwartz inequality 
\be
\alpha^{(J/\psi)} \, \alpha^{(\psi')} \ge |\alpha^{(\psi \psi')}|^2
\label{schw}
\ee
should hold. It can be expected however that each of the discussed diagonal parameters should exceed the off-diagonal one, and also that the chromo-polarizability for the $\pp$ and $\chi_c$ states is larger than that of the $\jp$ due to their larger spatial size.

The strength of the van der Waals type interaction between a charmonium state and a hadron (generically denoted here as $X$) can be evaluated using the effective Hamiltonian (\ref{hdiag}) and the well known expression for the conformal anomaly in QCD in the chiral limit:
\be
\theta_\mu^\mu = -{9 \over 32 \pi^2} \, F_{\mu \nu}^a F^{a \,\mu \nu}= {9 \over 16 \pi^2} \, \left ( E_i^a E_i^a - B_i^a B_i^a \right )~,
\label{anom}
\ee
where ${\vec B}^a$ is the chromo-magnetic field, and the (normalization) expression for the static average of the trace of the stress tensor $\theta_\mu^\mu$ over any state $X$ in terms of its mass $M_X$:
\be
\langle X | \theta_\mu^\mu ({\vec q}=0) | X \rangle = M_X~,
\label{mx}
\ee
which is written here assuming nonrelativistic normalization for the state $X$:
$\langle X | X \rangle=1$. Averaging the effective Hamiltonian (\ref{hdiag}) over a hadron $X$ made out of light quarks and/or gluons, one thus finds 
\be
\langle X | H_{\rm eff} | X \rangle \le - {8 \pi^2 \over 9} \, \alpha^{(\psi)} \, \, M_X~,
\label{ieff}
\ee
where the inequality arises from the fact that the average value of the full square operator $B_i^a B_i^a$ over a physical hadron $X$ has to be non-negative.
The relation (\ref{ieff}) shows an integral strength of the interaction. Namely, if the force between the charmonium ($\psi$) and the light hadronic matter inside the hadron $X$ is described by a potential $V({\vec x})$, such that $V$ goes to zero at large $|{\vec x}|$, this relation gives the bound for the integral
\be
\int V({\vec x}) \, d^3x \le - {8 \pi^2 \over 9} \, \alpha^{(\psi)} \, M_X~.
\label{iint}
\ee

The value of the integral in Eq.(\ref{iint}), although undoubtedly corresponding to an attraction, does not by itself automatically imply existence of a bound state, since it does not take into account the kinetic energy, which in a nonrelativistic treatment of the system is $p^2/2{\bar M} \sim 1/(R^2 \bar M)$, with $R$ being a characteristic size of the hadron $X$ and $\bar M = M_X M_\psi/(M_X+M_\psi)$ the reduced mass in the system. The spatial integral of the kinetic energy is then parametrically of order $R/{\bar M}$, and, given the relation in Eq.(\ref{iint}), the condition for existence of a bound state can be written as
\be
\alpha^{(\psi)} \, {M_X \, \bar M \over R} \ge C~,
\label{cond}
\ee
where $C$ is a numerical constant (parametrically of order one) which depends on a model for the distribution of the interaction over the interior of the hadron $X$ and thus on a more precise definition of the ``characteristic size" $R$.
Considering the condition (\ref{cond}) for excited light-matter resonances, one can readily see that the appearance of a hadro-charmonium state, possibly at a higher excitation, depends on the behavior of the combination $M_X \, \bar M / R$ with the excitation number of the resonance $X$. In particular, if the characteristic size $R$ grows slower than the mass $M_X$ a binding of charmonium necessarily occurs in a sufficiently highly excited resonance. Such behavior takes place, e.g. in the once popular bag model\cite{bagm}, where $R \propto M^{1/3}$. However in models which reproduce the approximately linear behavior of the Regge trajectories for the resonances, such as string model including its recently discussed\cite{kkss} implementation in terms of AdS/QCD correspondence with `linear' confinement, one effectively finds $R \propto M$, and a more accurate estimate of numerical factors in the condition (\ref{cond}) is generally needed, as well as a better knowledge of the chromo-polarizability for charmonium levels.

In lieu of a reliable theory, for an estimate of the numerical constant $C$ in Eq.(\ref{cond}), we approximate the interaction potential by a `square well' shape, i.e. $V(r)=-V_0$ at $r < R$ and $V(r)=0$ otherwise, and conservatively replace the inequality (\ref{iint}) by equality. Then the condition for existence of a bound state reads as
\be
\alpha^{(\psi)} \, {M_X \, \bar M \over R} \ge {3 \, \pi \over 16}~.
\label{condm}
\ee
At $\alpha^{(\psi)} = 2\,$GeV$^{-3}$ and $M_X=1\,$GeV (${\bar M} \approx 0.8\,$GeV) the latter criterion is satisfied at $R < 0.5\,$fm which certainly looks somewhat restrictive. Using instead $M_X=2\,$GeV, one would get a more encouraging condition $R < 1.8\,$fm. It should be emphasized however that the proper chromo-polarizability of at least the $\pp$ resonance is fully expected to exceed the `reference' value used in this estimate. Therefore we conclude that it is very likely that there exist bound states of charmonium with excited resonances of light hadronic matter in the range of the mass $M_X$ starting from below 2\,GeV.

If the observed resonance $Y(4.26)$ is interpreted as an `existence proof' of a hadro-charmonium state containing $\jp$, then one inevitably concludes that there should be similar states containing higher charmonium levels, in particular the $\chi_{cJ}$ states. Also hadro-charmonium resonances with spin-singlet charmonium levels, $\eta_c$, $\eta_c(2S)$ and $h_c$ should also exist, although it might be quite challenging to identify them experimentally. Furthermore, the applicability of our approximations and of the binding criterion (\ref{cond}) improves at larger mass $M_X$ of the host resonance. Therefore we expect existence of higher resonances decaying into specific charmonium states and light mesons, corresponding to higher excitation of the host light-matter resonance. The same inference can also be applied to baryonic resonances being hosts for hadro-charmonium: even if the binding of $\jp$ or of $\pp$ to a  nucleon is insufficient, it may become sufficient in higher baryonic resonances thus resulting in what can be called baryo-charmonium.

The interaction with gluonic field inside a light-matter hadron generally gives rise not only to an attraction of a fixed charmonium state, but also to transitions between different charmonium levels. The most obvious is the transition $\pp \to \jp$ in hadronic matter\cite{sv}, which would violate the so far discussed conservation of the quantum numbers of charmonium in a hadro-charmonium state.  The rate of such transition is determined by the off-diagonal chromo-polarizability, such as 
$\alpha^{(\psi \psi')}$. Another determining parameter is the form factor for the matrix element of the operator $E_i^a E_i^a$ over the light hadron state corresponding to the momentum transfer in such transition. Typically, the latter momentum is not small: e.g. $q^2 \approx 1.3\,$GeV$^2$ for a kinematically possible decay $Z(4.43) \to \pi \jp$. If the light-matter resonance in which the hadro-charmonium is formed has an extended characteristic size, the form factor can considerably suppress the rate of the charmonium cross-level transitions. The actual estimate of the rate of such processes is quite model-dependent, and can be very approximately evaluated as being of order of few MeV\cite{sv}. This is a relatively small fraction of the observed widths of the $Y$ and $Z$ peaks, however one very likely should expect existence of decays like $Z(4.43) \to \pi \jp$, or $Y(4.66) \to \pi \pi \jp$ with measurable branching ratios.

It should be also noted that in the previous discussion it was assumed that neither charmonium, nor the host light-matter resonance are significantly distorted by the van der Waals type interaction. This approximation holds when the size scales of the charmonium and the host resonance are very much different. In reality however the ratio of the scales may be only 2 $\div$ 3, so that a certain deformation of both the light and the charmonium states by the interaction should be expected. For the charmonium the smallness of the deformation is quantified by the discussed relative suppression of the transitions between its levels inside the system. For the host light-matter resonance, however, we have no means at present to quantify its deformation by the charmonium `kernel'.  Thus it is not clear at the moment how and to what extent the  hadro-charmonium states can be correlated with the actual known light-matter resonances.

The discussed picture of hadro-charmonium naturally invites a question about existence of similar states for bottomonium which is heavier and therefore the reduced mass $\bar M$ in a hadro-bottomonium state saturates at higher value. The problem with bottomonium is that its characteristic size is still smaller than that of charmonium: its known transitional chromo-polarizability $|\alpha^{(\Upsilon \Upsilon')}| \approx 0.6\,$GeV$^{-3}$~\cite{mvpol} is about three times smaller than the analogous quantity $\alpha^{(\psi \psi')}$ in charmonium. If this factor also indicates the corresponding reduction of diagonal chromo-polarizability for $\Upsilon$ and $\Upsilon(2S)$, the binding criterion (\ref{cond}) would require a higher mass of the light-matter resonance $M_X$, at which mass the decay width can become too large for the resonance to be identifiable experimentally. On the other hand, excited bottomonium states $\chi_b(2P)$, $\Upsilon(3S)$, $\Upsilon(1D)$ have a larger spatial size and may form hadro-bottomonium states at a lower mass $M_X$. The total mass of such states should be about 1\,GeV above the corresponding bottomonium levels, i.e. in the approximate range 11.0 - 11.5\,GeV. The signature of such state with e.g. the $\Upsilon(3S)$ bottomonium `core' would clearly be that it dominantly decays into $\Upsilon(3S)$ and pions (one or two - depending on the isospin), but not into channels with other bottomonium resonances or with $B$ meson pairs.

To summarize. We suggest that some of the recently found charmonium-related peaks, specifically the $Y$ and $Z$ resonances above 4.2\,GeV, decaying into a particular charmonium state and one or two pions, have a structure where that particular charmonium resonance is bound as a compact object inside an excited state of light hadronic matter. We refer to such systems as hadro-charmonium. We therefore expect that there should exist a rich spectrum of such states, containing other compact charmonium levels, in particular resonances decaying into $\chi_{cJ}$ and one or two pions, and also resonances where charmonium is bound inside still higher excitations of the light hadronic matter, including baryonic resonances. The transitions between charmonium states inside hadro-charmonium are expected to be relatively suppressed, but should show up at presumably measurable level as decays into channels with `non-preferred' charmonium state. Finally, similar systems containing bottomonium, if exist, are more likely to contain excited bottomonium levels, $\chi_{bJ}(2P)$, $\Upsilon(3S)$ and $\Upsilon(1D)$, which thus provides a signature for experimental search of hadro-bottomonium.

This work is supported in part by the DOE grant DE-FG02-94ER40823.


\end{document}